\documentclass[prd,twocolumn,superscriptaddress,floatfix,nofootinbib]{revtex4-2}
\pdfoutput=1 
\usepackage[T1]{fontenc}
\usepackage{amsmath}
\usepackage{amssymb}
\usepackage{amsthm}
\usepackage{amsfonts}
\usepackage{graphicx}
\usepackage{enumitem}
\usepackage{bm}
\usepackage{rotating}
\usepackage[colorlinks]{hyperref}
 \hypersetup{
     colorlinks=true,
     linkcolor=BlueViolet,
     filecolor=blue,
     citecolor=BrickRed, 
     urlcolor=cyan,
     }
\usepackage{pbox}
\usepackage[dvipsnames,usenames]{xcolor}
\usepackage{array}
\usepackage{physics}
\usepackage{dsfont}
\usepackage{mathtools}
\usepackage{leftidx}
\usepackage{lmodern}
\usepackage[normalem]{ulem}
\usepackage{braket}
\usepackage{tensor}
\usepackage{mathrsfs}
\usepackage{float}
\usepackage{dsfont}
\usepackage[margin=2cm]{geometry}
\usepackage[title]{appendix}
\usepackage{tcolorbox}
\usepackage{tocloft}
\usepackage{titlesec}
\usepackage{tikz}
\usetikzlibrary{arrows.meta,decorations.pathmorphing,math}

\renewcommand{\Re}{\mathrm{Re}}

\renewcommand{\tilde}{\widetilde}

\newcommand{\ii}{\mathsf{i}}

\newcommand{\sx}{\mathsf{x}}

\newcommand{\rr}[1]{\left(#1\right)}
\newcommand{\bx}{{\bm{x}}}

\newcommand{\bk}{{\bm{k}}}

\newcommand{\R}{\mathbb{R}}

\newcommand{\M}{\mathcal{M}}

\begin{document}

\title{Unruh phenomena and thermalization for qudit detectors}

\author{Caroline Lima}
\email{clima@perimeterinstitute.ca}
\affiliation{Department of Physics and Astronomy, University of Waterloo, Waterloo, Ontario, N2L 3G1, Canada}
\affiliation{Institute for Quantum Computing, University of Waterloo, Waterloo, Ontario, N2L 3G1, Canada}
\affiliation{Perimeter Institute for Theoretical Physics, 31 Caroline Street North, Waterloo, Ontario, N2L 2Y5, Canada}

\author{Everett Patterson}
\email{e2patte@uwaterloo.ca}
\affiliation{Department of Physics and Astronomy, University of Waterloo, Waterloo, Ontario, N2L 3G1, Canada}
\affiliation{Institute for Quantum Computing, University of Waterloo, Waterloo, Ontario, N2L 3G1, Canada}

\author{Erickson Tjoa}
\email{erickson.tjoa@mpq.mpg.de}
\affiliation{Max-Planck-Institut f\"ur Quantenoptik, Hans-Kopfermann-Stra\ss e 1, D-85748 Garching, Germany}
\affiliation{Department of Physics and Astronomy, University of Waterloo, Waterloo, Ontario, N2L 3G1, Canada}
\affiliation{Institute for Quantum Computing, University of Waterloo, Waterloo, Ontario, N2L 3G1, Canada}

\author{Robert B. Mann}
\email{rbmann@uwaterloo.ca}
\affiliation{Department of Physics and Astronomy, University of Waterloo, Waterloo, Ontario, N2L 3G1, Canada}
\affiliation{Institute for Quantum Computing, University of Waterloo, Waterloo, Ontario, N2L 3G1, Canada}
\affiliation{Perimeter Institute for Theoretical Physics, 31 Caroline Street North, Waterloo, Ontario, N2L 2Y5, Canada}

\date{\today}

\begin{abstract}

We study  Unruh phenomena for a qudit detector coupled to a quantized scalar field, comparing its response to that of a standard qubit-based Unruh-DeWitt detector. We show that there are limitations to the utility of the detailed balance condition as an indicator for Unruh thermality of higher-dimensional qudit detector models. This can be traced to the fact that a qudit has multiple possible transition channels between its energy levels, in contrast to the 2-level qubit model.
We illustrate these limitations using two types of qutrit detector models based on the spin-1 representations of $SU(2)$ and the non-Hermitian generalization of the Pauli observables (the Heisenberg-Weyl operators).

\end{abstract}

\maketitle

\section{Introduction}

The Unruh effect is one of the most remarkable predictions in quantum field theory: it says that accelerating observers do not perceive the Minkowski vacuum state as empty but rather as a thermal bath with temperature proportional to the proper acceleration of the observer \cite{Unruh1979evaporation}. It can be equivalently stated as the fact that the bifurcate Killing horizon of Rindler observers with proper acceleration $a$ can be assigned a temperature---the Unruh temperature $\mathsf{T}_{U} = a/(2\pi)$. At the level of quantum field theory, the Unruh effect is the statement that the pullback of the Wightman two-point functions with respect to the Minkowski vacuum state along a constant-acceleration trajectory is stationary and (anti-)periodic in  the imaginary time direction, with period equal to the inverse Unruh temperature $\beta = \mathsf{T}_{U}^{-1}$---this is an example of the \textit{Kubo-Martin-Schwinger} (KMS) \textit{condition}. It has been argued that the effect persists even for interacting theories \cite{unruh1984whathappens}. 

From the perspective of relativistic quantum information (RQI), the Unruh effect can be viewed as the statement that a uniformly accelerating  two-level system (``qubit'') interacting with a quantum field initialized in the Minkowski vacuum state thermalizes to a Gibbs state with temperature equal to the Unruh temperature. These results rely on the fact that the problem can be formulated ``quantum-optically'' using the so-called \textit{Unruh-DeWitt} (UDW) particle detector model \cite{Aubry2014derivative,Satz2006howoften,Benito2019asymptotic,Moustos2017nonmarkov,kaplanek2020hot,Fewster2016wait}. Sometimes this is understood as the  \textit{detailed balance condition} \cite{Pipo2019without}, which says that the excitation-to-deexcitation ratio (EDR)
of the detector 
is equal to $\exp(-\beta\Omega)$, where $\Omega$ is the energy gap of the detector. The detailed balance condition exploits the KMS condition and under some mild technical assumptions is a necessary and sufficient condition of thermalization for initial states that have no coherence in the Hamiltonian eigenbasis of its free Hamiltonian. Consequently, the detailed balance condition is often taken as a diagnostic for thermalization  (see also \cite{perche2021thermalization,tjoa2022unruhdewitt} for some generalization on this front).

In this paper, we are interested in a more thorough study of the Unruh effect using a generalization of the UDW detector model where the detector is a three-level system (``qutrit'') or higher. There are at least three reasons why this  generalization merits investigation:
\begin{enumerate}[leftmargin=*]
    \item In many situations, qubits exhibit certain coincidences that the higher-dimensional qudits or harmonic oscillators \cite{Brown2013harmonic,bruno2021measurement} do not share. For example, in a model where a qubit interacts with a bosonic environment (as is the case for Unruh phenomena),  at leading order in perturbation theory all qubit states that are diagonal in the energy eigenbasis cannot generate coherence. This will not be the case for higher-dimensional qudits, which in turn has implications on how we deal with the ultraviolet (UV) behavior of the environment.

    \item More importantly, unlike qubits, which  are essentially uniquely defined through their free Hamiltonian $\mathfrak{h}\sim \Omega \hat{n}\cdot\hat{\Vec{\sigma}}$ (for some energy gap $\Omega$, $\hat{n}$ a unit vector and $\hat{\vec{\sigma}}\equiv (\hat{\sigma}_x,\hat{\sigma}_y,\hat{\sigma}_z)$),  there are multiple inequivalent definitions of a $d$-level quantum system depending on the allowed transitions. For example, the spin-$1$ representation of an $SU(2)$ qutrit does not have the same internal dynamics as the qutrit constructed as a defining representation of $SU(3)$, or using the Heisenberg-Weyl operators.

    \item Qudit detector models have not been investigated much in the RQI context, with some rare exceptions (such as \cite{verdon2016asymptotic}). Most studies consideer two extremes, employing either qubit detector models or harmonic oscillator detectors. A more complete understanding of the various  possibilities would allow for extending known relativistic quantum information protocols to higher-dimensional detectors.

\end{enumerate}

We are particularly interested in the definition, meaning, and mechanism of thermalization for qutrits.
If the Unruh effect is to be taken operationally as a generic thermalization of a quantum-mechanical system moving along a uniformly accelerated trajectory, it is necessary that we understand the circumstances under which  thermalization of an accelerated detector occurs for generic detector models. Note that this generalization is distinct from the one studied in \cite{perche2021thermalization}, where the dynamics of the generalized detector model was restricted to a two-dimensional subspace.

Here we analyze the role of the detailed balance condition as a diagnostic/indicator for thermalization due to the Unruh effect. We show that there are strong limitations placed on the value of the detailed balance condition when we allow for higher-dimensional detector models. This can be traced to the fact that there are many possible types of three-level systems with different kinds of allowed transitions and degeneracies, while there is only essentially one type of qubit detector.  We map out these limitations by constructing two types of qudit detector model, based on the spin-$j$ representations of  $SU(2)$ and the non-Hermitian generalization of the Pauli observables of the qubit detector (the Heisenberg-Weyl operators \cite{Bertlmann2008,Vourdas2004}). We provide some connections to the detailed balance property associated with the Fourier transform of the Wightman functions.

Our paper is organized as follows. In Sec.~\ref{sec: general-setup} we establish the general setup for the physical situation we will analyze. In Sec. \ref{sec: SU2qudit} we study the $SU(2)$ qutrit detector and infer some properties of a higher dimensional $SU(2)$ qudit. In 
Sec. \ref{sec: heisenbergweyl}, we analyze the Heisenberg-Weyl qutrit detector model. 
In Sec. \ref{sec: conclusion} we discuss the general results and provide some future directions. In Appendix~\ref{appendix: integrals} we include technical details of our calculations and in Appendix~\ref{appendix:generalcaseSU2} we present the more general expression for our calculation in Sec. \ref{sec:SU(2)-qutrit} We use natural units $c=\hbar=1$ and the mostly-plus signature for the metric and we write $\sx$ to denote spacetime events.

\section{General setup}
\label{sec: general-setup}

In this section we give the general construction for the detector-field interaction needed to study Unruh phenomena. We will then specialize to some natural choice of detector-field coupling.

\subsection{Scalar field theory in Minkowski spacetime}
Let $\M$ be an $(n+1)$-dimensional Minkowski spacetime and consider a real scalar field $\phi$ obeying the Klein-Gordon equation
\begin{align}
    (\partial_\mu\partial^\mu-m^2){\phi} = 0\,.
\end{align}
Quantization gives rise to the scalar field operator $\hat\phi(\sx)$ that defines an operator-valued distribution: it can be expressed as a mode decomposition
\begin{align}
    \hat{\phi}(\sx) =\int\dd^n\bk\,\left[\hat{a}_\bk^{\phantom{\dagger}} u^{\phantom{*}}_\bk(\sx) + \hat{a}_\bk^\dagger u^*_\bk(\sx)\right] \,,
\end{align}
where $\{u_\bk(\sx)\}$ are the positive-frequency modes, and the ladder operators satisfy the canonical commutation relation $[\hat{a}^{\phantom{\dagger}}_{\bk},\hat{a}_{\bk'}^{\dagger}]= \delta^n(\bk-\bk')$. For the Unruh effect, we are interested in the quantization with respect to the global inertial frame associated with Minkowski coordinates $\sx\equiv (t,\bx)$. The corresponding ground state from this quantization is the \textit{Minkowski vacuum} $\ket{0_\textsc{M}}$ associated with the plane-wave mode
\begin{align}
    u_\bk(\sx) &= \frac{1}{\sqrt{2(2\pi)^n
    \omega_\bk}}e^{-\ii \omega_\bk t+\ii\bk\cdot\bx}\,,
\end{align}
such that $\hat{a}_\bk^{\phantom{\dagger}}\ket{0_\textsc{M}}=0$ for all $\bk$, $\omega_\bk = \sqrt{|\bk|^2+m^2}$. The modes are normalized with respect to the Klein-Gordon inner product
\begin{align}
    (f,g)_{\textsc{KG}} \coloneqq \ii\int_{\Sigma_t} \dd^n\bx\, \rr{f\partial_t g^* - g^*\partial_t f}\,,
\end{align}
so that
\begin{align}
    (u_\bk^{\phantom{*}},u_{\bk'}^{\phantom{*}})_{\textsc{kg}} &= \delta^n(\bk-\bk')\,, \;\;
    (u_\bk^*,u_{\bk'}^*)_{\textsc{kg}} = -\delta^n(\bk-\bk')\,,\notag \\
    (u_\bk^{\phantom{*}},u_{\bk'}^*)_{\textsc{kg}} &= 0\,.
\end{align}

We will be interested in two states:  the Minkowski vacuum state $\hat{\rho}_\textsc{M}^{\phantom{*}} \coloneqq \ketbra{0}{0}_\textsc{M}$ and the thermal state\footnote{Strictly speaking, in the Hilbert space built from the Minkowski vacuum, the thermal state cannot be written in terms of the density matrix $\hat{\rho}_\beta$. Hence the density matrix $\hat{\rho}_\beta$ should be understood as formal expression, but the Wightman two-point functions are always well-defined through the Kubo-Martin-Schwinger (KMS) construction --- see \cite{fewster2019algebraic} for details. } $\hat{\rho}_\beta$. These two states are examples of \textit{quasifree states} --- states that are completely characterized by their Wightman two-point functions, 
\begin{align}
    \mathsf{W}(\sx,\sx') &\coloneqq \braket{\hat{\phi}(\sx)\hat{\phi}(\sx')}_{\hat{\rho}} = \Tr\Bigr(\hat{\rho}\hat{\phi}(\sx)\hat{\phi}(\sx')\Bigr)\,.
\end{align}
For the vacuum state, the Wightman two-point function  is given by
\begin{align}
    \mathsf{W}_\textsc{M}(\sx,\sx') &= \int\frac{\dd^n\bk}{2(2\pi)^n\omega_\bk}e^{-\ii\omega_\bk(t-t')+\ii\bk\cdot(\bx-\bx')}\,,
\end{align}
where this is to be understood as a bi-distribution. This can be written as the closed-form expression  
\begin{align}
    \mathsf{W}_\textsc{M}(\sx,\sx') &= \lim_{\epsilon\to 0^+}\frac{m^{\frac{n-1}{2}}}{(2\pi)^{\frac{n+1}{2}}}\frac{1}{\rr{-(\Delta t-\ii\epsilon)^2+|\Delta\bx|^2}^{\frac{n-1}{4}}}\notag\\
    &\times K_{\frac{n-1}{2}}(m\sqrt{-(\Delta t-\ii\epsilon)^2+|\Delta\bx|^2})\,,
\end{align}
where $\Delta t = t-t',\Delta\bx = \bx-\bx'$ and $K_\alpha(z)$ is the modified Bessel function of the second kind of order $\alpha$. For the thermal state, it can be shown that (see, e.g., \cite{simidzija2018harvesting})
\begin{align}
    \mathsf{W}_\beta(\sx,\sx') &= \mathsf{W}_\textsc{M}(\sx,\sx') + \mathsf{W}_{\beta,\text{reg}}(\sx,\sx') \,,\notag\\
    \mathsf{W}_{\beta,\text{reg}}(\sx,\sx') &\coloneqq \int \frac{\dd^n\bk}{2(2\pi)\omega_\bk}\frac{e^{-\ii\omega_\bk(t-t')+\ii\bk\cdot(\bx-\bx')}+\text{c.c.}}{e^{\beta\omega_\bk}-1}\,,
    \label{eq: thermal bath-full}
\end{align}
where $\beta = \mathsf{T}_{U}^{-1}$ is the inverse Unruh temperature. The splitting of $\mathsf{W}_\beta$ into the singular (distributional) vacuum piece $\mathsf{W}_\textsc{M}$ and the regular thermal piece $\mathsf{W}_{\beta,\text{reg}}$ follows directly from the fact that all physically reasonable states are Hadamard states \cite{wald1994quantum,KayWald1991theorems}. This observation will be of importance in our subsequent analysis.

\subsection{Detector-field coupling for a pointlike accelerating detector in vacuum}

Consider a \textit{pointlike} qudit detector moving along an accelerated trajectory  $\sx(\tau) \equiv (t(\tau),x(\tau),\bx_\perp(\tau))$ parametrized by proper time $\tau$, with
\begin{align}
    t(\tau) &= \frac{1}{a}\sinh a\tau\,,\quad x(\tau) = \frac{1}{a}\cosh a\tau\,,
    \label{eq: trajectory-accel}
\end{align}
and $\bx_\perp(\tau) = \bm{0}$. The parameter $a$ is the proper acceleration along the trajectory. For studying the Unruh phenomenon, we prescribe the following interaction Hamiltonian (in the interaction picture):
\begin{align}
    \hat{H}_I(\tau) &= \lambda\chi(\tau)\hat{O}(\tau)\otimes\hat{\phi}(\sx(\tau))\,,
    \label{eq: interaction-hamiltonian}
\end{align}
where $\hat{O}$ is a Hermitian observable of the detector. The time-dependent operator $\hat{O}(\tau)$ is obtained from free evolution via the free Hamiltonian of the system $\mathfrak{h}$:
\begin{align}
    \hat{O}(\tau) &= e^{\ii\hat{\mathfrak{h}}\tau} \hat{O} e^{-\ii\hat{\mathfrak{h}} \tau}\,.
\end{align}
The usual UDW detector model \cite{Unruh1979evaporation,DeWitt1979} corresponds to a qubit detector with
\begin{align}
    \hat{\mathfrak{h}} = \frac{\Omega}{2}(\hat{\sigma}^z+\openone)\,,\qquad \hat{O} = \hat{\sigma}^x = \ketbra{e}{g} + \ketbra{g}{e}\,,
\end{align}
where $\ket{g},\,\ket{e}$ are the eigenstates of the free Hamiltonian with energy $0,\,\Omega$ respectively. 

Since we are considering qudit detector models, the natural form of the interaction Hamiltonian is given by \eqref{eq: interaction-hamiltonian}, with different detector models corresponding to different specifications of $\hat{\mathfrak{h}}$ and $\hat{O}$. 
The unitary time evolution for a generic interaction \eqref{eq: interaction-hamiltonian} is given by
\begin{align}
    \hat{U} &= \mathcal{T}\exp\left[-\ii\int_{-\infty}^\infty \dd\tau\,\hat{H}_I(\tau)\right]\,.
\end{align}
If the joint detector-field state is initially prepared in the uncorrelated state $\hat{\rho}_{-\infty} = \hat{\rho}_{\textsc{d},-\infty}\otimes\hat{\rho}_{\phi,-\infty}$,  the final state of the joint system is given by
\begin{align}
    \hat{\rho}_{\infty} &= \hat{U}\hat{\rho}_{-\infty}\hat{U}^\dagger\,. 
\end{align}
Perturbatively, up to second order in $\lambda$, we have
\begin{subequations}
\begin{align}
    \hat{U} &= \openone + \hat{U}^{(1)} + \hat{U}^{(2)} + \mathcal{O}(\lambda^3)\,, \\ 
    \hat{U}^{(1)} &= -\ii\int_{-\infty}^\infty \dd\tau\,\hat{H}_I(\tau)\,,\\
    \hat{U}^{(2)} &= -\int_{-\infty}^\infty\dd\tau \int_{-\infty}^\tau\dd\tau'\,\hat{H}_I(\tau)\hat{H}_I(\tau')\,,
\end{align}
\end{subequations}
where $\hat{U}^{(j)}$ are corrections of order $\lambda^j$.  

The final state of the detector can be obtained by tracing out the field's degrees of freedom. However, since the quasifree states have vanishing odd-point functions,  the leading-order perturbative corrections occur at $\lambda^2$ and corrections with odd powers of $\lambda$ are absent. Hence we write
\begin{align}
    \hat{\rho}_{\textsc{d},\infty} &= \hat{\rho}_{\textsc{d},-\infty} + \hat{\rho}^{(2)} + \mathcal{O}(\lambda^4)\,,
    \label{eq: final-state-quasifree}
\end{align}
where $\hat{\rho}^{(2)}=\sum_{k+l=2}\hat{\rho}^{(k,l)}$ includes all the perturbative corrections of order $\lambda^2$, with each $\hat{\rho}^{(k,l)}$ defined to be (traceless) perturbative corrections to the density matrix of the detector of order $
\lambda^{k+l}$:
\begin{equation}
    \hspace{-0.15cm}
    \begin{aligned}
    \hat{\rho}^{(k,l)} \coloneqq \tr_\phi(\hat{U}^{(k)}\hat{\rho}_{\textsc{d},-\infty}\hat{U}^{(l)\dagger})\,.
    \end{aligned}
\end{equation}
The perturbative correction to the detector's density matrix depends on the \textit{pullback} of the Wightman two-point function along the trajectory $\mathsf{\sx}(\tau)$, denoted $\mathsf{W}(\tau,\tau')\equiv \mathsf{W}(\sx(\tau),\sx(\tau'))$. For an accelerating detector with constant proper acceleration $a$, we have \cite{Takagi1986noise}
\begin{align}
    \mathsf{W}_a(\tau,\tau') &= \frac{1}{(2\pi)^{\frac{n-1}{2}}}
    \rr{\frac{m}{z_\epsilon}}^{\frac{n-3}{2}}K_{\frac{n-3}{2}}(mz_\epsilon)\,,
\end{align}
where 
\begin{align}
    z_\epsilon \equiv z_{\epsilon}(\tau-\tau') &= \frac{2\ii}{a}\sinh\rr{\frac{a}{2}(\tau-\tau'-\ii\epsilon)}\,.
\end{align}
For a massless scalar field ($m=0$)  the dispersion relation $\omega_\bk = \sqrt{|\bk|^2+m^2}$ reduces to $\omega_\bk = |\bk|$, and the Wightman two-point function simplifies to
\begin{align}
    \mathsf{W}_a(\tau,\tau') &= \frac{\Gamma\rr{\frac{n-3}{2}}}{4\pi^{\frac{n-1}{2}}}\frac{1}{z_\epsilon^{n-1}}\,,
\end{align}
where $\Gamma(z)$ is the Gamma function.  For simplicity we will henceforth specialize to the massless scalar field in $(3+1)$-dimensional Minkowski spacetime that is commonly studied in the literature. Since the Wightman function is \textit{stationary}, i.e., $\mathsf{W}_a(\tau,\tau') = \mathsf{W}_a(\tau-\tau')$, we can write 
\begin{align}
    \mathsf{W}_a(u) &= -\frac{a^2}{16\pi^2}\frac{1}{\sinh^2\rr{\frac{a}{2}(u-\ii\epsilon)}}\,,
    \label{eq: pullback-accel}
\end{align}
where $u=\tau-\tau'$.

\subsection{Thermalization of particle detectors}

The Unruh phenomenon can also be viewed as follows: the pullback of the vacuum Wightman two-point function along the accelerated trajectory \eqref{eq: pullback-accel} is equal to the pullback of the \textit{thermal} Wightman two-point function associated with an inertial trajectory $\sx(\tau) = (\tau,\bm{0})$, which reads
\begin{align}
    \mathsf{W}_\beta(u) &= -\frac{1}{4\beta^{2}}\frac{1}{\sinh^2\rr{\frac{\pi}{\beta}(u-\ii\epsilon)}}\,.
    \label{eq: thermal bath-pointlike}
\end{align}
Therefore an accelerated observer experiences thermal excitation even in the (Minkowski) vacuum environment, with Unruh temperature given by $\mathsf{T}_{U}\coloneqq \beta^{-1} = a/(2\pi)$. How do we make precise such a statement in an operational manner?

In general, to say that a qudit detector thermalizes to the Unruh temperature means that the final state of the detector approaches a steady state that is a Gibbs state, i.e., 
\begin{align}
    \lim_{T\to\infty} \hat{\rho}_{\textsc{d},\infty} &= \frac{e^{-\beta\hat{\mathfrak{h}}}}{\tr e^{-\beta\hat{\mathfrak{h}}}}\,,
\end{align}
where $\mathfrak{h}$ is the free Hamiltonian of the detector\footnote{This should also mean, implicitly, that we are in the weak coupling regime \cite{anders2022meanforce}.} and $T$ is the effective duration of interaction. However, to prove that this steady state is achieved is tricky for several reasons.

First of all, thermalization of the detector should be independent of its initial state. Consider for instance  a qubit detector model interacting with a thermal environment: in the limit of switching for adiabatically long times\footnote{The Gaussian switching guarantees this limit to be adiabatic; in the usual approach where sharp switching is used, it is necessary that the coupling strength is ``weakened'' at long times \cite{Garay2016anti-unruh,Pipo2019without}.} $T\to\infty$, the \textit{excitation-to-deexcitation} (EDR) \textit{ratio} satisfies the \textit{detailed balance condition} 
\begin{align}
    \lim_{T\to\infty} \frac{\Pr(\Omega)}{\Pr(-\Omega)} &= e^{-\beta\Omega}\,,
    \label{eq: qubit-EDR}
\end{align}
but this is not sufficient unless one proves that the off-diagonal terms in the energy eigenbasis vanish at long interaction times. That said, the detailed balance condition is often taken as an indicator of thermalization \cite{Pipo2019without,Fewster2016wait}; alternatively, one can typically show using e.g., master equations in open system dynamics, that at late times the detectors do have vanishing coherences under some conditions \cite{kaplanek2020hot,Moustos2017nonmarkov,Benito2019asymptotic}. Clearly, the Gibbs state obeys the detailed balance condition but extended to all energy levels:
\begin{align}
    \frac{\Pr(E_i\to E_j)}{\Pr(E_j\to E_i)} = e^{-\beta(E_j-E_i)}\,,\quad E_j \geq E_i\,.
    \label{eq: qudit-EDR}
\end{align}
Thus we expect that if the qudit detector thermalizes, then at the very least Eq.~\eqref{eq: qudit-EDR} should hold for sufficiently long interaction times and all coherences should decay appropriately.

In the next section we will study how the Unruh phenomenon is captured in the qudit generalization of the UDW detector model and study the value of the detailed balance condition in these models.

\section{Unruh effect for SU(2) qudit detector models}
\label{sec: SU2qudit}

The interaction for the $SU(2)$-qudit detector is obtained by replacing the monopole operator for the spin-1/2 qubit $\hat{\sigma}_x$ with the generic angular momentum operator $\hat{J}_x$ for spin-$j$ qudit representation of $SU(2)$ with $d=2j+1$. For the accelerated pointlike qudit model, we can write the interaction Hamiltonian density as
\begin{align}
    \hat{H}_I(\tau) &= \lambda \chi(\tau)\hat{J}_x(\tau)\otimes \hat{\phi}(\sx(\tau))\,,
    \label{eq: qudit-interaction-accel}
\end{align}
where $\sx(\tau)$ is given by \eqref{eq: trajectory-accel}. The free Hamiltonian of the detector can be taken to be
\begin{align}
    \hat{\mathfrak{h}} &= \Omega(\hat{J}_z+ j \openone)\,,
    \label{eq: qudit-free}
\end{align}
where $j$ is the orbital angular momentum number. The energy eigenstates are therefore naturally expressed in terms of the Dicke basis $\{\ket{j,m}: m = -j,-j+1,...,j-1,j\}$, with energy spectrum $\{0,\Omega,...,(2j-1)\Omega, 2j\Omega\}$. As before, the identity operator is just a global shift to set the ground state energy equal to zero.  The final state of the qudit detector is computed according to Eq.~\eqref{eq: final-state-quasifree}.

\subsection{SU(2) qutrit detectors}
\label{sec:SU(2)-qutrit}

For concreteness, in what follows we focus on a spin-1 qutrit $(j=1)$;  we will discuss the generalization later. We will use the shorthand $\ket{m}\equiv\ket{j=1,m}$.  In the usual Dicke basis ordered as $\{\ket{1},\ket{0},\ket{-1}\}$, we have
\begin{align}
    J_x = \frac{1}{\sqrt{2}}
    \begin{bmatrix}
    0 & 1 & 0 \\ 1 & 0 & 1 \\ 0 & 1 & 0
    \end{bmatrix}\,,\qquad 
    J_z = \begin{bmatrix}
    1 & 0 & 0 \\ 0 & 0 & 0 \\ 0 & 0 & -1
    \end{bmatrix}\,.
\end{align}

Suppose the detector is initially in a diagonal state
\begin{align}
    \hat{\rho}_{\textsc{d},-\infty}
    &= \begin{bmatrix}
        a & 0 & 0 \\ 0 & b & 0 \\ 0 & 0 & c
    \end{bmatrix}\,,\quad a+b+c=1\,.
\end{align}
Our task is to calculate the leading-order perturbative corrections $\hat{\rho}^{(j,k)}$ for $j+k=2$. We have
\begin{widetext}
\begin{subequations}
    \begin{align}
    \hat{\rho}^{(1,1)} 
    &\equiv \frac{\lambda^2}{2} \int\dd t\,\dd t' \chi(t)\chi(t') \mathsf{W}(t,t')
    \begin{bmatrix}
    b e^{-\ii \Omega  (t-t')} & 0 & b e^{\ii \Omega  (t+t')} \\
    0 &  a e^{\ii \Omega (t-t')}+ c e^{-\ii \Omega (t-t')} & 0 \\
     b e^{-\ii \Omega  (t+t')} & 0 & b e^{\ii \Omega  (t-t')} \\
    \end{bmatrix}\,,\\
    \rho^{(2,0)} 
    &= -\frac{\lambda^2}{2}\int\dd t \dd t' \Theta(t-t')\chi(t)\chi(t') \mathsf{W}(t,t')
    \begin{bmatrix}
     a e^{\ii \Omega(t-t')}  & 0 & c e^{\ii \Omega (t+t')} \\
     0 & b \left( e^{\ii \Omega(t-  t')}+ e^{-\ii \Omega(t-t')} \right) & 0 \\
     a e^{-\ii \Omega (t+t')} & 0 & c e^{-\ii \Omega(t- t')}
    \end{bmatrix}\,,\\
    \rho^{(0,2)} &= \rho^{(2,0)\dagger}\,.
\end{align}
\end{subequations}
\end{widetext}

By defining the following integrals:
\begin{equation}
    \begin{aligned}
    \mathcal{I} &\coloneqq \lambda^2 \int\dd t\,\dd t'\,\chi(t)\chi(t')e^{\pm \ii\Omega(t+t')}\mathsf{W}(t,t')\,,\\
    \mathcal{L}_{\pm} &\coloneqq \lambda^2 \int\dd t\,\dd t'\,\chi(t)\chi(t')e^{\pm\ii\Omega (t-t')}\mathsf{W}(t,t')\,,\\
    \mathcal{Q} &\coloneqq \lambda^2 \int\dd t\,\dd t'\,\Theta(t-t')\chi(t)\chi(t')e^{\pm\ii\Omega(t+t')}\mathsf{W}(t,t')\,, \\
    \mathcal{R}_\pm &\coloneqq \lambda^2 \int\dd t\,\dd t'\,\Theta(t-t')\chi(t)\chi(t')e^{\pm\ii\Omega(t-t')}\mathsf{W}(t,t')\,,
    \end{aligned}
    \label{eq: integrals}
\end{equation}
the final state takes the form of an $X$-state 
\begin{align}
    \hat{\rho}_{\textsc{d},\infty}
    &=
    \begin{bmatrix}
        a + \rho_{11}^{(2)} & 0 & \rho_{13}^{(2)} \\ 0 & b+\rho_{22}^{(2)} & 0 \\ \rho_{13}^{(2)*} & 0 & c+\rho_{33}^{(2)}
    \end{bmatrix}+\mathcal{O}(\lambda^4)\,,
\end{align}
where 

\begin{equation}
\begin{aligned}
    \rho_{11}^{(2)} &= \frac{1}{2}\Bigr(b\mathcal{L}_- - \, a\mathcal{L}_+\Bigr)\,,\\
    \rho_{22}^{(2)} &= \frac{1}{2} \Bigr(a\mathcal{L}_+ + c\mathcal{L}_- -\,b(\mathcal{L}_-+\mathcal{L}_+)\Bigr)\,,\\
    \rho_{33}^{(2)} &= \frac{1}{2}\Bigr(b\mathcal{L}_+ -\, c\mathcal{L}_-\Bigr)\,,\\
    \rho_{13}^{(2)} &= \frac{1}{2}\Bigr(b\mathcal{I} - a\mathcal{Q}^* - c\mathcal{Q}\Bigr)\,.
\end{aligned}
\end{equation}
Observe that unlike the qubit detector model, starting from initially diagonal state generically produces nonzero coherence.

At this stage, if we were to interpret the detailed balance condition  as taking 
place in the adiabatic long-time limit $T\to\infty$, we would also need to show that  $\rho_{13}^{(2)}\to 0$ in the long-time limit.  By direct computation via the Sokhotsky formula, we have (see Appendix~\ref{appendix: integrals})

\begin{widetext}
\begin{subequations}
\begin{align}
    \mathcal{I} &= \lambda^2 e^{-\frac{1}{2} T^2 \Omega ^2} \left(
    \frac{1}{4\pi}+ \sqrt{2\pi}aT\int_{-\infty}^\infty \dd u\,e^{-2u^2/(aT)^2} \frac{\left(1- u^2 \text{csch}^2u\right)}{16 \pi ^2 u^2}\right)\,,
    \\
    \mathcal{Q} &=\frac{\lambda^2}{2} e^{-\frac{1}{2} T^2 \Omega ^2}
    \rr{ \frac{1}{4 \pi }-\frac{\ii T^3}{4 \pi  \epsilon\left({\epsilon}^2+ T^2\right)} + \sqrt{2\pi}aT\int_{0}^\infty \dd u\,e^{-2u^2/(aT)^2} \frac{\left(1- u^2 \text{csch}^2u\right)}{16 \pi ^2 u^2}}\,,
\end{align}
\end{subequations}
\end{widetext}
where $\epsilon>0$ is a fixed UV cutoff.
From these expressions it can be shown for fixed $\epsilon,\Omega > 0$ that
\begin{align}
    \lim_{T\to\infty} \mathcal{I} = \lim_{T\to\infty} \mathcal{Q} = 0
    \label{eq: Q-decay}
\end{align}
for any fixed $\Omega,a>0$. Note that unlike the qubit scenario, even for an initially diagonal state the time evolution gives rise to a UV divergence contained in  the integral $\mathcal{Q}$  for the pointlike model. For our purposes, we have used a UV regulator $\epsilon$ by smoothing the Dirac delta function by its Gaussian nascent delta family (\textit{c.f.} Appendix~\ref{appendix: integrals}). However, the physical prediction is largely independent of the UV divergences so long as we are in the sufficiently late time regime --- that is, for fixed $\Omega,\epsilon$ such that $\Omega T \gg 1$ and $aT\gtrsim 1$. Therefore, the coherence term does not obstruct us from relying on the detailed balance condition for probing thermalization.

For the qubit scenario, the EDR is umambiguously defined since the diagonal entries only has one free parameter (due to the unit trace condition). However, the qutrit case is slightly trickier.  
For the $SU(2)$ qutrit, we have $\ket{-1},\ket{0},\ket{1}$ as the energy eigenstates, and
we might expect   the EDR to have the detailed balance property
\begin{equation} 
    \begin{aligned}
    \frac{\Pr(E_{-1}\to E_0)}{\Pr(E_0\to E_{-1})}&=\frac{\Pr(E_{0}\to E_1)}{\Pr(E_1\to E_{0})}\stackrel{T\to\infty}{=} e^{-\beta\Omega}\,,\\
    \frac{\Pr(E_{-1}\to E_1)}{\Pr(E_1\to E_{-1})} &\stackrel{T\to\infty}{=} e^{-2\beta\Omega}\,.
    \end{aligned}
\end{equation}
The small subtle
distinction between the qubit and qutrit cases
has to do with the well-known fact that the EDR is defined for two \textit{different initial conditions}. Classically, the detailed balance condition is defined by comparing two processes in which one  is the reverse  of the other: in this case, we have the de-excitation process as a \textit{reverse} of the excitation process \cite{crooks1998nonequilibrium,jarzynski2000hamiltonian,aw2021bayesian}. Following this strategy, an analogous construction for qutrits can be done as follows. For the $\ket{-1} \to \ket{0}$ transition, we set $b=c=0,\; a=1$ and for $\ket{0}\to\ket{-1}$ we set $a=c=0,\; b=1$ and  we get
\begin{align}
    \lim_{T\to\infty }\frac{\Pr(E_{-1}\to E_0)}{\Pr(E_0\to E_{-1})} &= \lim_{T\to\infty }\frac{\mathcal{L}_-}{\mathcal{L}_+} \to e^{-\beta\Omega}\,,
\end{align}
which agrees with the qubit scenario since $\mathcal{L}_\pm$ are precisely the transition probabilities for the qubit detector with energy gap $\Omega$. Similarly, for the $\ket{0} \to \ket{1}$ transition, we set $a=c=0,\; b=1$ and for $\ket{1}\to\ket{0}$ we set $a=b=0,\; c=1$ obtaining
\begin{align}
    \lim_{T\to\infty }\frac{\Pr(E_{0}\to E_1)}{\Pr(E_1\to E_{0})} &= \lim_{T\to\infty }\frac{\mathcal{L}_-}{\mathcal{L}_+}  = e^{-\beta\Omega}\,.
\end{align}
This again agrees with the qubit scenario.

The last case, associated with the transition $\ket{-1}\leftrightarrow \ket{1}$, is subtle: for the $\ket{-1} \to \ket{1}$ transition, we need to set $b=c=0,\; a=1$ and for $\ket{1}\to\ket{-1}$ we need  $a=b=0,\; c=1$. However, for this qutrit model, at leading order in perturbation theory we have 
\begin{align}
    \Pr(E_{-1}\to E_1) \,,  \Pr(E_{1}\to E_{-1}) \sim \mathcal{O}(\lambda^4)\,,
\end{align}
so at the level of this computation the detailed balance condition between $\ket{\pm 1}$ is \textit{indeterminate}. From the perspective of the model, this problem arises because the $SU(2)$ qutrit model does not allow for a ``direct'' transition from $\ket{-1}\leftrightarrow\ket{1}$ with a single application of the monopole operator $\hat{J}_x$; consequently, the detailed balance can only hold at the level of the detector dynamics for the ``nearest-neighbor'' energy level allowed by the monopole operator. This is to be contrasted with the generalized qubit model in \cite{perche2021thermalization}, where the monopole operator allows for any 
(fixed) two-level subspace of a qudit detector. 

It is reasonable, in light of the above results, to posit   that the detailed balance condition makes sense in some restricted way by excluding the $\ket{-1}\leftrightarrow \ket{1}$ transitions, i.e., the detailed balance condition is still valid in the ``qubit subspace'' where the direct transition via $\hat{J}_x$ is allowed   at leading order in perturbation theory.
However, as we will show next, once the initial state contains non-vanishing coherence, on its own the detailed balance condition is not a useful diagnostic for thermalization within the standard Dyson perturbative expansion without further assumptions or approximations.

When working with an $SU(2)$ qutrit, it is convenient to decompose the initial density matrix into two distinct subspaces that we call the $X$-block and the $O$-block in the energy eigenbasis, i.e.,
\begin{align}
    \hat{\rho}_{-\infty} &= \hat{\rho}_{-\infty,X} + \hat{\rho}_{-\infty, O}\,,\notag\\ 
    \hat{\rho}_{-\infty, X} &= \begin{bmatrix}
        \rho_{11} & 0 & \rho_{13} \\
        0 & \rho_{22}  & 0 \\
        \rho_{31} & 0  & \rho_{33} 
    \end{bmatrix},\, \hat{\rho}_{-\infty, O} = \begin{bmatrix}
        0 & \rho_{12} & 0 \\
         \rho_{21} & 0  & \rho_{23} \\
        0 & \rho_{32}  & 0 
    \end{bmatrix}.
\end{align}
This decomposition is convenient when the field is in a quasifree state since, due to the vanishing of odd-point functions of the field, each block will evolve independently in the sense that if we initialize the state in an $X$-state (with zero component in the $O$-block), the final state will remain in the $X$-block subspace. 

Now consider instead the following initial state \footnote{The expression for the second-order corrections for general initial states are given in Appendix \ref{appendix:generalcaseSU2}.} 
\begin{align}
    \ket{\psi}=& \frac{1}{\sqrt{2}} (\ket{1}+\ket{0}), 
    \notag 
    \end{align}
    so that
    \begin{align}
    \hat{\rho}_{\textsc{d},-\infty} =& \ketbra{\psi}{\psi} = \frac{1}{2} \begin{pmatrix}
        1 & 1 & 0\\
        1 & 1 & 0\\
        0 & 0 & 0
    \end{pmatrix}.
\end{align}
Then, using the integrals \eqref{eq: integrals} the leading-order corrections that are second-order in $\lambda$ are
\begin{align}
 \hat{\rho}^{(2)} &=
 \frac{1}{4} \begin{bmatrix}
    \mathcal{L}_- - \mathcal{L}_+
    & \mathcal{I} -\mathcal{L}_+ - \mathcal{R}_-^*
    & \mathcal{I} - \mathcal{Q}^* \\
    \mathcal{I}- \mathcal{L}_+-\mathcal{R}_-
    &  - \mathcal{L}_-
    &  \mathcal{L}_+ - \mathcal{Q}^* \\
  \mathcal{I} - \mathcal{Q}
  &  \mathcal{L}_+  - \mathcal{Q} 
  &  \mathcal{L}_+ \\
 \end{bmatrix}.
\end{align}
Notice that in the long time limit, some of the corrections of the decoherence terms in the final state are not only non-zero but are also divergent in the limit $T\gg \infty$. To see this, observe that at fixed $\Omega$ and $a$ we have 
\begin{align}
    \lim_{a T\gg 1}\mathcal{L}_\pm \propto \lambda^2 aT\,,
    \label{eq: secular}
\end{align}
so the (de-)excitation probabilities increase linearly with the duration of interaction $T$. Crucially, this means that for sufficiently long interaction times, the perturbative calculation   breaks down and becomes unreliable essentially due to the \textit{secular growth} (the phenomenon where $\lambda^2 aT$ viewed as a time-dependent coupling constant is unbounded as $T\to\infty$) \cite{kaplanek2020hot}.

More importantly, a moment's consideration would inform us that at this level of perturbative computation we will not be able to have a diagnostic of thermalization even if the detailed balance condition can be satisfied: by inspecting the coherence term $\rho_{12}$ of the \textit{full state}, we see that
\begin{align}
    \rho_{12} = \frac{1}{2} + \frac{1}{4}(\mathcal{I} -\mathcal{L}_+ - \mathcal{R}_-^*) \not\to 0
\end{align}
in the long-time limit $T\to\infty$ (which we argued is divergent in the naïve long-time limit).  The problem is threefold:
\begin{itemize}[nolistsep]
    \item[(i)] We do not have an analogous EDR-type statement for off-diagonal matrix elements. Therefore, the only clue to thermalization behavior for these components is to require them to vanish at late times as required by the Gibbs state.
    \item[(ii)] However, since the perturbative correction suffers from secular growth due to the presence of $\mathcal{L}_+$ (\textit{c.f.} Eq.~\eqref{eq: secular}), the late-time limit is not reliable at this stage of the computation. 
    \item[(iii)] From the nature of the perturbative calculation, it is impossible to reliably make the coherence term vanish since by construction the corrections are small. This is analogous to requiring that $e^{-x}\approx 1-x\to 0$, which is outside the validity of the approximations. 
\end{itemize}
From a physical perspective, we expect that the Unruh effect should properly thermalize the detector to the Unruh temperature $\mathsf{T}_{U}=a/(2\pi)$.  This naïve example demonstrates the difficulty of probing thermalization of even simple systems armed with only the bare Dyson series expansion and the detailed balance condition. It is also not hard to see that the same problem would arise for qubits the moment the initial state is allowed to have coherence: in this case, it is not automatic that   detailed balance is, on its own, sufficient for thermalization. 

In other investigations of the   qubit detector model, thermalization of an accelerating qubit detector interacting with a relativistic scalar bath are worked out using open master equations \cite{kaplanek2020hot,kaplanek2023effective,Benito2019asymptotic,Moustos2017nonmarkov}. There it can be shown that the detector thermalizes properly in the sense of approaching an appropriate Gibbs state, provided additional approximations and assumptions to avoid the secular growth are satisfied. Essentially, by restricting our attention to some ``high-temperature'' regimes, it is possible to perform late-time resummations of the second-order corrections \cite{kaplanek2020hot,kaplanek2023effective}, which is precisely what the master equations \cite{lindblad1975completely,gorini1978} are designed for. The point is that there are additional techniques \textit{within perturbation theory} that one can use to study thermalization due to the Unruh effect (or Hawking effect in the case of black holes).

\subsection{Some features of SU(2) qudit detectors}

The decomposition of the $SU(2)$ quantum state into $X$- and $O$-blocks generalizes to higher spin systems with recognizable patterns. Consider the spin-2 qudit, initialized in the $\ketbra{0}{0}$ state with respect to the ordered Dicke basis $\{\ket{2},\ket{1},\ket{0},\ket{-1},\ket{-2}\}$. The ensuing perturbed state is given by
\begin{widetext}
\begin{equation}
    \hat{\rho}_{\ketbra{0}{0},\infty} = \begin{bmatrix}
        0 & 0 & -\sqrt{\frac{3}{2}}\mathcal{Q}_{+} & 0 & 0 \\
        0 & \frac{3}{2}\mathcal{L}_{-} & 0 & \frac{3}{2}\mathcal{I}_{-} & 0 \\
        \sqrt{\frac{3}{2}}\mathcal{Q}_{+}^* & 0 & 1 -\frac{3}{2}(\mathcal{L}_{+}+\mathcal{L}_{-}) & 0 & -\sqrt{\frac{3}{2}}\mathcal{Q}_{-}^* \\
        0 & \frac{3}{2}\mathcal{I}_{+} & 0 & \frac{3}{2}\mathcal{L}_{+} & 0 \\
        0 & 0 & -\sqrt{\frac{3}{2}}\mathcal{Q}_{-} & 0 & 0 
    \end{bmatrix} + \mathcal{O}(\lambda^4)\,.
    \label{eq: ququint}
\end{equation}
\end{widetext}
Observe that within the second-order perturbative expansion, starting from the middle state $\ketbra{
0}{0}$ only allows us to perturb the state at most two entries away from the center of the density matrix \eqref{eq: ququint}. The reason is simple --- the Dyson series truncated at second order can only contain at most two products of the angular momentum operators $\hat{J}_x(\tau)\hat{J}_x(\tau')$; thus it can only map basis elements to at most $\ketbra{j}{k}\to \ketbra{j\pm \ell}{k\pm m}$, where $\ell+m=2$ are non-negative integers. Consequently, the detailed balance argument will only work well between nearest-neighbor energy levels using the same kind of arguments for the qutrit case. Note that in the above example, the coherences can be shown to vanish in the long-time regime, so again for a generic $SU(2)$ qudit detector model initialized in diagonal states, the detailed balance condition will still be a useful (partial) indicator for the thermalization due to Unruh effect as we have shown for the qutrit model.

We stress that thermalization is not a purely non-perturbative effect  just because the detailed balance condition alone does not suffice for diagnosing thermalization in the standard Dyson series approach. Indeed, the standard open quantum systems approach involves weak coupling regimes (see, e.g., \cite{breuer2002theory,lidar2020lecture}) that must be amenable to perturbative techniques. The real problem is that of extracting \textit{reliable} late-time predictions from perturbative calculations.  The master equation approach furnishes one way to do so. Certainly, the general problem of thermalization of (possibly strongly) interacting systems is both difficult and an active area of research  \cite{Anders2022meanforce,Gogolin2016closedequi,Gogolin2011thermalization,Linden2009thermal,Short2011thermal,Deutsch2018ETH}. 

Last, but not least, we  mention that in the context of the UDW model, the calculations we have done to diagnose thermalization are clearly model-dependent. In the qubit scenario, there is not much freedom in varying the model since  there is only one possible free Hamiltonian that describes  (non-degenerate) two-level systems and the qubit observables are very restrictive, i.e., there are no selection rules that forbid some direct transitions between energy levels. In contrast, once we consider qudit detectors with Hilbert space dimension of at least three, the story changes: there are several ways to define a three-level system and it is possible to design interaction Hamiltonians that forbid some direct transitions. In the $SU(2)$ model, we have the situation where the free Hamiltonian is non-degenerate with equal energy spacing and the direct transition between $\ket{\pm1}$ is not allowed. We will see in Section~\ref{sec: heisenbergweyl} that indeed some conclusions can change by allowing for other kinds of three-level systems.

\section{Heisenberg-Weyl qudit detector model and the value of detailed balance}
\label{sec: heisenbergweyl}

Let us briefly consider a different generalization of the qudit detector, using a different generalization of $\hat{\sigma}_x,\hat{\sigma}_z$ known as the \textit{clock and shift matrices}, defined by
\begin{align}
    \hat{X} &= \sum_{j=0}^{d-1}\ketbra{j+1\!\!\!\!\mod d}{j}\,,\quad 
    \hat{Z} = \sum_{j=0}^{d-1}e^{2\pi\ii j/d}\ketbra{j}{j}
    \,.
\end{align}
For simplicity we focus on the qutrit case, where the  clock and shift matrices are given by
\begin{align}
    \hat{X} &= \begin{bmatrix}
        0 & 0 & 1 \\
        1 & 0 & 0 \\
        0 & 1 & 0
    \end{bmatrix}\,,\quad 
    \hat{Z} = \begin{bmatrix}
        1 & 0 & 0 \\
        0 & e^{\frac{2\pi\ii}{3}} & 0 \\
        0 & 0 & e^{\frac{-2\pi\ii}{3}}
    \end{bmatrix} \,,
\end{align}
where we use the $\{\ket{0},\ket{1},\ket{2}\}$ basis. This generalization is \textit{not} Hermitian, so we propose that the Heisenberg-Weyl (HW) qudit model\footnote{This is to be compared with the model studied in \cite{verdon2016asymptotic} which constructs a different interaction Hamiltonian without using Hermitian observables for the detector. The model, however, has acausal coupling.} is given by
\begin{align}
    \hat{H}_I(\tau) &= \lambda\chi(\tau)\bigr(\hat{X}(\tau)+\hat{X}^\dagger(\tau)\bigr)\otimes\hat{\phi}(\sx(\tau))\,,
\end{align}
with free Hamiltonian $\hat{\mathfrak{h}} = \Omega(\hat{Z}+\hat{Z}^\dagger)/2$. 
The free Hamiltonian  has degenerate ground states:
\begin{align}
    \frac{\Omega}{2}(\hat{Z} + \hat{Z}^\dagger) = \frac{\Omega}{2} \begin{bmatrix}
        2 & 0 & 0 \\
        0 & -1 & 0 \\
        0 & 0 & -1
    \end{bmatrix} 
\end{align}
with $E_1-E_0=0$ and   energy gap   $E_2-E_0=E_2-E_1=\frac{3}{2}\Omega$.

To simplify our work with the Heisenberg-Weyl qudits, we must introduce two new integrals, in addition to making some slight modifications to our previous integral notation as follows: 

\begin{widetext}
\begin{subequations}
\label{eq: integral-variants-qudits}
\begin{align}
    \mathcal{L}_{q}  &\coloneqq \lambda^2 \int\dd t\,\dd t'\,\chi(t)\chi(t')e^{\ii q\Omega (t-t')}\mathsf{W}(t,t')\,,\\
    \mathcal{R}_{q} &\coloneqq \lambda^2 \int\dd t\,\dd t'\,\Theta(t-t')\chi(t)\chi(t')e^{\ii q\Omega(t-t')}\mathsf{W}(t,t')\,, \\
    \mathcal{U}_{q} &\coloneqq \lambda^2 \int\dd t\,\dd t'\,\chi(t)\chi(t')e^{\ii q\Omega t}\mathsf{W}(t,t')\,, \\
    \mathcal{V}_{q}^\pm &\coloneqq \lambda^2 \int\dd t\,\dd t'\,\Theta(\pm(t-t'))\chi(t)\chi(t')e^{\ii q\Omega t}\mathsf{W}(t,t')\,. 
\end{align}
\end{subequations}
\noindent where $q$ is a real number scaling the energy gap $\Omega$. 

Suppose that the detector is initialized in the diagonal state 
\begin{equation}
    \hat{\rho}_\text{D} = \begin{bmatrix}
        a & 0 & 0 \\
        0 & b & 0 \\
        0 & 0 & c
    \end{bmatrix},\qquad a+b+c=1\,.
\end{equation}
The second order corrections are given by 
\begin{align}
    \hat{\rho}_\text{D}^{(1,1)} &= a \begin{bmatrix}
        0 & 0 & 0 \\
        0 & \mathcal{L}_{+\frac32} & \mathcal{L}_{+\frac32}\\
        0 & \mathcal{L}_{+\frac32} & \mathcal{L}_{+\frac32} 
    \end{bmatrix} + b \begin{bmatrix}
        \mathcal{L}_{-\frac32} & 0 & \mathcal{U}_{+\frac32}^* \\
        0 & 0 & 0\\
        \mathcal{U}_{+\frac32} & 0 & \mathcal{U}_0
    \end{bmatrix} + c \begin{bmatrix}
        \mathcal{L}_{-\frac32} & \mathcal{U}_{+\frac32}^* & 0  \\
        \mathcal{U}_{+\frac32} & \mathcal{U}_0 & 0\\
        0 & 0 & 0
    \end{bmatrix}
\end{align}
and
\begin{align}
    \hat{\rho}_\text{D}^{(2,0)}+\hat{\rho}_\text{D}^{(0,2)} =& -a \begin{bmatrix}
        2 \mathcal{L}_{-\frac32} & \mathcal{V}_{+\frac32}^{-} & \mathcal{V}_{+\frac32}^{-} \\
        \mathcal{V}_{+\frac32}^{-*} & 0 & 0 \\
        \mathcal{V}_{+\frac32}^{-*} & 0 & 0
    \end{bmatrix} -b \begin{bmatrix}
        0 & \mathcal{V}_{+\frac32}^+ & 0 \\
        \mathcal{V}_{+\frac32}^{+*} & \mathcal{U}_0 + \mathcal{L}_{-\frac32} & \mathcal{R}_{-\frac32}^* \\
        0 & \mathcal{R}_{-\frac32} & 0
    \end{bmatrix} \nonumber \\ 
    & -c \begin{bmatrix}
        0 & 0 & \mathcal{V}_{+\frac32}^{+} \\
        0 & 0 & \mathcal{R}_{-\frac32} \\
        \mathcal{V}^{+*}_{+\frac32} & \mathcal{R}_{-\frac32}^* & \mathcal{U}_0 + \mathcal{L}_{-\frac32}
    \end{bmatrix}\,.
\end{align}
Following the same strategy as the $SU(2)$ qutrit and using  the notation $P_{i\to j}\equiv \Pr(E_i\to E_j)$, the EDRs for the HW qutrit are given by:
\begin{align}
    \frac{P_{1\to2}}{P_{2\to1}} &= \frac{\mathcal{L}_{-\frac32}}{\mathcal{L}_{+\frac32}} = e^{-\beta\frac32\Omega} \,,\quad 
    \frac{P_{0\to2}}{P_{2\to0}} = \frac{\mathcal{L}_{-\frac32}}{\mathcal{L}_{+\frac32}} = e^{-\beta\frac32\Omega} \,,\quad 
    \frac{P_{0\to1}}{P_{1\to0}} = \frac{\mathcal{U}_0}{\mathcal{U}_0} =1 = e^{-\beta(0)\Omega}\,.
\end{align}
\end{widetext}

Observe that for this qutrit detector model, the detailed balance condition effectively ``works''. 
The reason is quite straightforward: due to the nature of the shift operator $\hat{X}$, all the energy levels are essentially nearest-neighbors and therefore, the EDR captures properly all the forward and reverse processes. For exactly this reason, the HW qudits with Hilbert space dimension $\geq 4$ will \textit{not} have the detailed balance condition work for all transitions since the direct transition $\ket{0}\to\ket{2}$ is not allowed: some indeterminacy will appear analogously to the $SU(2)$ case.

The point of this brief analysis is to emphasize that the value of the detailed balance condition as an indicator for thermalization depends on the model, i.e., what we take as the detector and its free Hamiltonian, as well as the choice of coupling with the field observable. The latter affects the allowed transitions at leading order in perturbation theory and whether the detailed balance condition \eqref{eq: qudit-EDR} holds for all pairwise energy levels depends on both choices. If, however, our goal of using a detector to probe the field is to simply extract, say, the Unruh temperature, then in practice we do \textit{not} need the detailed balance condition to hold for all energy levels. Yet, in this case one is left with the question of whether the qudit coupled to the field truly thermalizes (even if the matrix elements contain information about the Unruh temperature).

\section{Discussion and outlook}
\label{sec: conclusion}

In this paper we analyzed the Unruh effect using $SU(2)$ and the Heisenberg-Weyl qutrit detector models. We also expanded our analysis to understand some general features of higher dimensional qudits in both models. We concluded that the detailed-balance condition, commonly taken as an indicator for thermalization of qubits, is not satisfied in general for higher dimensional systems. In fact, we observed that whenever there is a selection rule in the internal dynamics of the detector, the final state of the detector up to second order in perturbation theory will not be thermal. It will contain coherences that do not vanish in the long time limit.
We also noticed that the Heinsenberg-Weyl qutrit is a special case  for which the detailed-balance condition is satisfied. Indeed, because in this case there is no selection rule, i.e., all possible jumps between states are allowed, it behaves similarly to the qubit case.

All the considerations in this paper suggest that we should not require the detailed balance condition for the {probe} system (detector) to be the litmus test for the Unruh effect beyond the qubit model, unless one has good control over long-time regimes. When we couple a qubit detector model to probe the Unruh effect, what we are really trying to do is to probe the thermal behavior of the \textit{quantum field} as seen by an accelerating observer --- that is, we are trying to extract the properties of the \textit{pullback of the Wightman function} $\mathsf{W}_a(\tau,\tau') $. However, due to the stationary of the Wighman function with respect to the proper time $\tau$, by writing $u=\tau-\tau'$ we can compute the Fourier transform
\begin{align}
    \tilde{\mathsf{W}}_a(\omega) \coloneqq \int\dd u\, \mathsf{W}_a(u)e^{-\ii\omega u}\,,
\end{align}
and this obeys the relation \cite{Takagi1986noise,kaplanek2020hot}
\begin{align}
    \frac{\tilde{\mathsf{W}}_a(\omega)}{\tilde{\mathsf{W}}_a(-\omega)} = e^{-\beta\omega}\,,\quad \beta^{-1} = \frac{a}{2\pi}\,,
    \label{eq: EDR-QFT}
\end{align}
which, as a statement about power spectra, can be regarded as a  detailed balance relation that is a 
consequence of the KMS condition
\cite{Takagi1986noise}.

What the standard UDW detector model traditionally was designed to do is to show that with an appropriate choice of switching functions, coupling and energy gaps, it is possible to recover \eqref{eq: EDR-QFT} in appropriate limits \cite{Takagi1986noise,Satz2006howoften,Satz2007transitionrate,Aubry2014derivative,louko2016unruh,Fewster2016wait,Pipo2019without,tjoa2022unruhdewitt}. However, the full characterization of thermality of the field state is still given by the Kubo-Martin-Schwinger (KMS) conditions \cite{Kubo1957thermality,Martin-Schwinger1959thermality}.

In short, our work shows that in general, the only reliable way to probe thermal behavior of the field is to show that the probe thermalizes to the Gibbs state in appropriate limit, unless one restricts to only using two-level detectors. This is because only in the context of qubit detectors (or restricting to two-level subspaces of a qudit detector) that the lhs of Eq.~\eqref{eq: EDR-QFT} can be faithfully mapped to the EDR of the qubit detector, from which the detailed balance condition of the detector is equivalent in appropriate (adiabatic) limit to the detailed balance coming from the field-theoretic calculations.

Our work constitutes one of the few investigations in to higher-dimensional detector models in the context of RQI, and so  several future directions arise naturally from our analysis. For example, the well-known entanglement harvesting protocol \cite{pozas2015harvesting,tjoa2020harvesting} remains largely unexplored in higher dimensional system. This is not surprising, since there is a lack of good measures of entanglement beyond negativity and concurrence for mixed states of two qudits (with local dimension $\geq 3$). We leave these questions for the future.

\section*{Acknowledgment}

 E. T. acknowledges funding from the Munich Center for Quantum Science and Technology (MCQST), funded by the Deutsche Forschungsgemeinschaft (DFG) under Germany’s Excellence Strategy (EXC2111 - 390814868). E. P. acknowledges the funding from Ontario Graduate Scholarship. The authors thank the participants of RQI-N 2023 for helpful comments. This work was supported in part by the Natural Sciences and Engineering Research Council of Canada. This work was conducted on the traditional territory of the Neutral, Anishnaabeg, and Haudenosaunee Peoples; the University of Waterloo and the Institute for Quantum Computing are situated on the Haldimand Tract, land that was promised to Six Nations. Research at Perimeter Institute is supported in part by the Government of Canada through the Department of Innovation, Science and Economic Development Canada and by the Province of Ontario through the Ministry of Colleges and Universities.

\newpage
\appendix
\begin{widetext}

\section{Integral computations for accelerating detector}
\label{appendix: integrals}

In this Appendix we compute the following integrals that appear in the matrix elements of the detector:
\begin{subequations}
    \begin{align}
    \mathcal{I} &\coloneqq \lambda^2\int\dd \tau\,\dd \tau'\,\chi(\tau)\chi(\tau')e^{\pm \ii\Omega(\tau+\tau')}\mathsf{W}_a(\tau,\tau')\,,\\
    \mathcal{L}_{\pm} &\coloneqq \lambda^2\int\dd \tau\,\dd \tau'\,\chi(\tau)\chi(\tau')e^{\pm\ii\Omega (\tau-\tau')}\mathsf{W}_a(\tau,\tau')\,,\\
    \mathcal{Q}  &\coloneqq \lambda^2\int\dd \tau\,\dd \tau'\,\Theta(\tau-\tau')\chi(\tau)\chi(\tau')e^{\pm\ii\Omega(\tau+\tau')}\mathsf{W}_a(\tau,\tau')\,, \\
    \mathcal{R}_\pm &\coloneqq  \lambda^2\int\dd \tau\,\dd \tau'\,\Theta(\tau-\tau')\chi(\tau)\chi(\tau')e^{\pm\ii\Omega(\tau-\tau')}\mathsf{W}_a(\tau,\tau')\,,
    \end{align}
\end{subequations}
where the pullback of the Wightman function along the accelerated trajectory is given by
\begin{align}
    \mathsf{W}_a(\tau,\tau') = -\frac{a^2}{16\pi^2}\frac{1}{\sinh^2(\frac{a}{2}(\tau-\tau'-\ii\epsilon))} \equiv \mathsf{W}_a(\tau - \tau')\,.
\end{align}
The Wightman function is stationary with respect to the proper time $\tau$ since it is a function of $\tau-\tau'$. It will be very useful to formally split the Wightman function into two pieces, the singular piece that carries the distributional singularities and the regular piece that is a proper function (and not a tempered distribution). The way to do this is to expand around $a=0$, so that in fact the singular contribution is given by the pullback of the vacuum Wightman function along the \textit{inertial trajectory}: 
\begin{align}
    \mathsf{W}_a(\tau,\tau') &= \mathsf{W}_{\textsc{M}}(\tau,\tau') + \mathsf{W}_{a,\text{reg}}(\tau,\tau')\,,\qquad \mathsf{W}_\textsc{M}(\tau,\tau') = -\frac{1}{4\pi^2}\frac{1}{(\tau-\tau'-\ii\epsilon)^2}\,.
    \label{eq: hadamard-split}
\end{align}
In what follows we will also make extensive use of the following change of variable:
\begin{align}
    u = t-t'\,,\qquad v=t+t'\,,\qquad \dd \tau\,\dd\tau' = \frac{1}{2}\dd u\,\dd v\,.
    \label{eq: null-coordinates}
\end{align}
We will also use Gaussian switching function $\chi(\tau) = e^{-\tau^2/T^2}$. Finally, a very useful tool for us is the \textbf{Sokhotsky formula}:
\begin{align}
    \frac{1}{(s\pm \ii\epsilon)^n} &= \text{p.v.}\frac{1}{s^n}\pm \frac{(-1)^n}{(n-1)!}\ii\pi\delta^{(n-1)}(s)\,,
    \label{eq: Sokhotsky}
\end{align}
where p.v. denotes Cauchy principal value and $\delta^{(n)}$ denotes the $n$-th weak derivative of the Dirac delta distribution. The general strategies here would also work for variants of the above integrals when one considers higher-dimensional qudits (e.g., the ones in Eq.~\eqref{eq: integral-variants-qudits}).

\noindent \paragraph*{\textbf{Computation of $\mathcal{I}$.}} 

Using the change of variable \eqref{eq: null-coordinates}, we have
\begin{align}
    \mathcal{I} &= \frac{1}{2}\lambda^2\int\dd u\,\dd v\, \chi\!\left(\frac{v+u}{2}\right)\chi\!\left(\frac{v-u}{2}\right)e^{\pm\ii \Omega v}\mathsf{W}_a(u)\notag\\
    &= \frac{1}{2}\lambda^2\int\dd u\,\dd v\, \chi(u/\sqrt{2})\chi(v/\sqrt{2})e^{\pm\ii \Omega v}\mathsf{W}_a(u)\notag\\
    &= \frac{1}{2}\lambda^2\int\dd v\,\chi(v/\sqrt{2})e^{\pm\ii \Omega v}\int\dd u \,\chi(u/\sqrt{2})\mathsf{W}_a(u) \notag\\
    &= \lambda^2\sqrt{\frac{\pi}{2}}T e^{-\frac{T^2\Omega^2}{2}}\int\dd u \,\chi(u/\sqrt{2})\mathsf{W}_a(u)\,.
\end{align}
In the second equality we have used the fact that the switching is Gaussian and observe that the $\pm$ sign in the phase does not matter, hence we write $\mathcal{I}$ instead of $\mathcal{I}_\pm$. Next, using the splitting \eqref{eq: hadamard-split}, we have
\begin{align}
    \int\dd u \,\chi(u/\sqrt{2})\mathsf{W}_a(u) &= \int\dd u \,\chi(u/\sqrt{2})\mathsf{W}_\textsc{M}(u) + \int\dd u \,\chi(u/\sqrt{2})\mathsf{W}_{a,\text{reg}}(u)\,.
\end{align}
The second term corresponds to the finite-acceleration term that comes from the regular piece $\mathsf{W}_{a,\text{reg}}$ and vanishes in the limit $a\to 0$. It turns out that the regular piece does not admit a closed-form expression but it is straightforward to calculate numerically due to the non-distributional nature of the integrand. The first term can be computed exactly via the Sokhotsky formula\footnote{We can also evaluate it by rewriting the Wightman function $\mathsf{W}_\textsc{M}(\tau,\tau')$ in momentum space, i.e., as a distributional integral
\begin{align*}
    \mathsf{W}_\textsc{M}(\tau,\tau') &= \int\frac{\dd^n\bk}{2(2\pi)^n|\bk|}e^{-\ii|\bk|(\tau-\tau')}\,.
\end{align*}}. Together, we get
\begin{align}
    \mathcal{I} &= \lambda^2e^{-\frac{T^2 \Omega ^2}{2} } \left(
    \frac{1}{4\pi}+ \sqrt{2\pi}aT\int_{-\infty}^\infty \dd s\,e^{-2s^2/(aT)^2} \frac{\left(1- s^2 \text{csch}^2s\right)}{16 \pi ^2 s^2}\right) \,.
    \label{eq: I-integral}
\end{align}

\paragraph*{\noindent \textbf{Computation of $\mathcal{L}_\pm$.}} First, it is worth noting that $\mathcal{L}_\pm$ corresponds exactly to the transition probabilities for the qubit detector in the standard UDW model. In our convention, $\mathcal{L}_-$ is the {excitation} probability from the ground state, while $\mathcal{L}_+$ is the de-excitation probability. The calculation proceeds almost identically to the one for $\mathcal{I}$ except for the phase:
\begin{align}
    \mathcal{L}_\pm 
    &= \frac{1}{2}\lambda^2\int\dd u\,\dd v\, \chi(u/\sqrt{2})\chi(v/\sqrt{2})e^{\pm\ii \Omega u}\mathsf{W}_a(u)\notag\\
    &= \frac{1}{2}\lambda^2\int\dd v\,\chi(v/\sqrt{2})\int\dd u \,\chi(u/\sqrt{2})e^{\pm\ii \Omega u}\mathsf{W}_a(u) \notag\\
    &= \lambda^2\sqrt{\frac{\pi}{2}}T\int\dd u \,\chi(u/\sqrt{2})e^{\pm\ii \Omega u}\mathsf{W}_a(u)\,.
\end{align}
The integral over $u$ can be viewed as the Fourier transform of $\chi(u/\sqrt{2})\mathsf{W}_a(u)$. Using the splitting \eqref{eq: hadamard-split}, we get
\begin{align}
    \mathcal{L}_\pm &= 
    \underbrace{\frac{\lambda^2}{4\pi}\left(e^{-\frac{T^2\Omega^2}{2}} \pm \sqrt{\frac{\pi}{2}}\Omega T\text{erfc}\!\left(\tfrac{\mp \Omega T}{\sqrt{2}}\right)\right)}_{\text{vacuum contribution}}+\frac{\lambda^2a T}{4\sqrt{2\pi^3}}\int^\infty_0\dd s\frac{\cos(2\Omega s/a)e^{ - 2s^2/(aT)^2}(\sinh^2s-s^2)}{s^2\sinh^2s}\,.
\end{align}
The value for $\mathcal{L}_-$ is equal to $\mathcal{L}_{jj}/\lambda^2$ calculated in \cite{GallockYoshimura2023accel}. Note that only the vacuum contribution is sensitive to the $\pm$ sign while the regular finite-acceleration piece is symmetric under the exchange $\Omega\to -\Omega$. The regular part can also be shown to vanish as $a\to 0$.\\

\noindent\paragraph*{\textbf{Computation of $\mathcal{Q}$.}} 

Using the change of variable \eqref{eq: null-coordinates}, we have
\begin{align}
    \mathcal{Q}  
    &= \frac{1}{2}\lambda^2\int\dd u\,\dd v\, \Theta(u)\chi(u/\sqrt{2})\chi(v/\sqrt{2})e^{\pm\ii \Omega v}\mathsf{W}_a(u)\notag\\
    &= \frac{1}{2}\lambda^2\int\dd v\,\chi(v/\sqrt{2})e^{\pm\ii \Omega v}\int\dd u \,\Theta(u)\chi(u/\sqrt{2})\mathsf{W}_a(u) \notag\\
    &= \lambda^2\sqrt{\frac{\pi}{2}}T e^{-\frac{T^2\Omega^2}{2}}\int\dd u \,\chi(u/\sqrt{2})\Theta(u)\mathsf{W}_a(u)\,.
\end{align}
Again, using the split \eqref{eq: hadamard-split},
\begin{align}
    \mathcal{Q}_{\textsc{M}} = \lambda^2\sqrt{\frac{\pi}{2}}T e^{-\frac{T^2\Omega^2}{2}}\int\dd u \,\chi(u/\sqrt{2})\Theta(u)\mathsf{W}_\textsc{M}(u)\,,\quad \mathcal{Q}_{a,\text{reg}} = \lambda^2\sqrt{\frac{\pi}{2}}T e^{-\frac{T^2\Omega^2}{2}}\int\dd u \,\chi(u/\sqrt{2})\Theta(u)\mathsf{W}_{a,\text{reg}}(u)\,,
\end{align}
so that $\mathcal{Q} = \mathcal{Q}_{\textsc{M}}+\mathcal{Q}_{a,\text{reg}}$. Observe from Eq.~\eqref{eq: I-integral} that the regular piece is an integral over a symmetric function:
\begin{align}
    \mathcal{I}_{a,\text{reg}}\coloneqq \lambda^2e^{-\frac{T^2 \Omega ^2}{2} } \sqrt{2\pi}aT\int_{-\infty}^\infty \dd s\,e^{-2s^2/(aT)^2} \frac{\left(1- s^2 \text{csch}^2s\right)}{16 \pi ^2 s^2}\,.
\end{align}
Therefore, the Heaviside function in $\mathcal{Q}_{a,\text{reg}}$ cuts the integral by half, so we have
\begin{align}
    \mathcal{Q}_{a,\text{reg}} = \frac{1}{2}\mathcal{I}_{a,\text{reg}}\,.
\end{align}
The vacuum piece $\mathcal{Q}_{\textsc{M}}$, however, is problematic because it is UV-divergent: this is due to the common distributional singularity along $\tau-\tau'=0$. 
In order to regularize the UV singularity, let us first use the Sokhotsky formula and better flesh out the singularity. From the Sokhotsky formula, we have
\begin{align}
    \mathsf{Q}_{\textsc{M}} &\coloneqq -\frac{1}{4\pi^2}\int\dd u\,\Theta(u) e^{-u^2/(2T^2)}\rr{\frac{1}{u^2} - \ii\pi \delta^{(1)}(u)} \notag\\
    &= -\frac{1}{4\pi^2}\frac{1}{4 \sqrt{2\pi^3 T^2}} - \frac{\ii}{4\pi}\int\dd s\,\Theta(u)e^{-u^2/{2T^2}}\delta^{(1)}(u)\,.
\end{align}
In the second equality, the first term can be calculated in two ways, either via the Fourier transform or by direct integration over $\R\setminus [0,a)$ and subtraction of the divergent contribution. The second term is the origin of the singular behavior as it contains products of Dirac delta functions. To see this, we can define the inner product between some distribution $f$ and a  test function $g$ of $f$ such that
\begin{align}
    \braket{f,g} \coloneqq \text{p.v.}\int_{-\infty}^\infty \dd u\, f(u)^*g(u)\,.
\end{align}
Noting that test functions must have vanishing support at the boundaries (as they are defined to have compact supports), this inner product has the property that $\braket{f',g} = -\braket{f,g'}$. In particular, the definition of weak derivative of Dirac delta distribution is given in terms of the inner product as
\begin{align}
    \braket{\delta^{(n)},g} = (-1)^n{g^{(n)}}(0) \equiv (-1)^n\frac{\dd^n g}{\dd u^n}(0)\,.
\end{align}
Applying these to $\Theta$ and $\delta^{(1)}$, we have $\braket{\Theta,\delta^{(1)}} = -\braket{\Theta^{(1)},\delta} = -\braket{\delta,\delta}$ and we know that $\braket{\delta,\delta}$ is divergent. This is the statement that $\delta^{(1)}$ is not in the space of test functions of the Heaviside function (viewing Heaviside as a distribution), and vice versa.

UV regularization essentially resolves the issue above by putting a cutoff and forcing the integral defined through the inner product to be finite. There are at least two natural ways we can apply the UV regularization:
\begin{enumerate}[leftmargin=*,label=(\arabic*)]
    \item Regularize the delta function in the Sokhotsky formula, e.g., using Gaussian nascent family
    \begin{align}
        \delta_{a_0}(x) \coloneqq \frac{1}{\sqrt{2\pi a_0^2}}e^{-x^2/(2a_0^2)} \,.
    \end{align}
    We can either compute the nascent derivative $\delta_{a_0}^{(1)}$ or use the inner product: we get
    \begin{align}
        \mathsf{Q}_{\textsc{M}} &=-\frac{1}{4\pi^2}\frac{1}{4 \sqrt{2\pi^3 T^2}}-\frac{\ii T^2}{\sqrt{32\pi^3}a_0 \left({a_0}^2+ T^2\right)}\,.
    \end{align}

    \item Regularize the Heaviside function, e.g., using tanh:
    \begin{align}
        \Theta_{a_0}(x) \coloneqq \frac{1+\tanh(x/a_0)}{2}\,.
        \label{eq: approx-step}
    \end{align}
    In this case we get after using the inner product formula for distributions
    \begin{align}
        \mathsf{Q}_{\textsc{M}} &= -\frac{1}{4\pi^2}\frac{1}{4 \sqrt{2\pi^3 T^2}}-\frac{\ii}{8 \pi a_0}\,.
    \end{align}
\end{enumerate}
Hence, depending the choice of the UV regulator we see that
\begin{align}
    \mathcal{Q}_{\textsc{M}} &= \begin{cases}
    \displaystyle
    \lambda^2\left(\frac{e^{-\frac{1}{2} T^2 \Omega ^2}}{8 \pi }-\frac{\ii T^3 e^{-\frac{1}{2} T^2 \Omega ^2}}{8 \pi  a_0\left({a_0}^2+ T^2\right)}\right) \qquad \qquad &\text{(Method 1)}\\
    \displaystyle\\
    \displaystyle
    \lambda^2\left(\frac{e^{-\frac{1}{2} T^2 \Omega ^2}}{8 \pi }-\frac{\ii T e^{-\frac{1}{2} T^2 \Omega ^2}}{8 \sqrt{2 \pi } {a_0}} \right)
    &\text{(Method 2)}
    \end{cases}
\end{align}
For our purposes, the important feature of the regularization procedure is that in the long time regime $T\gg a_0$, the two methods give the same scaling behavior $\mathsf{Q}_{\textsc{M}}\sim -\ii/a_0$. Furthermore, the $a_0$-dependent term is very strongly suppressed at large $T\Omega\gg 1$ for fixed $a_0$ (in units of $\Omega$).

\noindent\paragraph*{\textbf{Computation of $\mathcal{R}_\pm$.}} 

Using the change of variable \eqref{eq: null-coordinates}, we have
\begin{align}
    \mathcal{R}_\pm  
    &= \frac{1}{2}\lambda^2\int\dd v\,\chi(v/\sqrt{2})\int\dd u \,\Theta(u)\chi(u/\sqrt{2})\mathsf{W}_a(u) \notag\\
    &= \lambda^2\sqrt{\frac{\pi}{2}}T \int\dd u \,\chi(u/\sqrt{2})\Theta(u)e^{\pm\ii \Omega u}\mathsf{W}_a(u)\,.
\end{align}
Regularizing $\mathcal{R}_\pm$ is less straightforward than the rest of the integrals due to the extra phase factor. Again, using the splitting \eqref{eq: hadamard-split}, let us write
\begin{align}
    \mathcal{R}_{\pm ,\textsc{M}} = \lambda^2\sqrt{\frac{\pi}{2}}T \int\dd u \,\chi(u/\sqrt{2})\Theta(u)e^{\pm \ii\Omega u}\mathsf{W}_\textsc{M}(u)\,,\quad \mathcal{R}_{\pm a,\text{reg}} = \lambda^2\sqrt{\frac{\pi}{2}}T \int\dd u \,\chi(u/\sqrt{2})\Theta(u)e^{\pm \ii\Omega u}\mathsf{W}_{a,\text{reg}}(u)\,,
\end{align}
so that $\mathcal{R} = \mathcal{R}_{\pm \textsc{M}}+\mathcal{R}_{\pm a,\text{reg}}$. The regular piece is straightforward and follows the same steps as $\mathcal{Q}_{a,\text{reg}}$: 
\begin{align}
    \mathcal{R}_{\pm a,\text{reg}} &=  \sqrt{2\pi}\lambda^2aT\int_{0}^\infty \dd s\, e^{\pm2\ii\Omega s/a} e^{-2s^2/(aT)^2}\frac{\left(1- s^2 \text{csch}^2s\right)}{16 \pi ^2 s^2}\,.
\end{align}
The singular part is much less straightforward. However, instead we can also compute this numerically by using the approximate Heaviside step function \eqref{eq: approx-step}: that is, we compute instead
\begin{align}
     \mathcal{R}_{\pm ,\textsc{M}} \to  \lambda^2\sqrt{\frac{\pi}{2}}T \int\dd u \,\chi(u/\sqrt{2})\Theta_{a_0}(u)e^{\pm \ii\Omega u}\mathsf{W}_\textsc{M}(u)\,.
\end{align}
For any finite $a_0>0$, one can show that the real part is well-behaved and the UV divergence is completely contained in the imaginary part (analogous to the UV divergence in $\mathcal{Q}_\text{M}$).

\section{Final state of the SU(2)-qutrit detector initialized in the most general state} \label{appendix:generalcaseSU2}
Suppose the detector is initialized in the state
\begin{align}
    \hat{\rho}_{\textsc{d},-\infty}
    &= \begin{bmatrix}
        a & d & e \\ d^* & b & f \\ e^* & f^* & c
    \end{bmatrix}\,,
\end{align}
where $a+b+c=1$. Following the same notation as for Section \ref{sec: SU2qudit}, we find for the leading-order corrections:
\begin{align}
 \hat{\rho}^{(2)} &=
 \frac{1}{2} \begin{bmatrix}
    -a \mathcal{L}_{+} +b\mathcal{L}_{-} -2\Re[e \mathcal{Q}^*]
    & d^*\mathcal{I} -d (\mathcal{L}_{+}+\mathcal{R}_{-}^*) +f \mathcal{L}_{-} -f^*\mathcal{Q} 
    & -a \mathcal{Q}^* +b \mathcal{I} -c \mathcal{Q} -e (\mathcal{R}_{-}^*+ \mathcal{R}_{+}) \\
    d\mathcal{I} -d^* (\mathcal{L}_{+}+\mathcal{R}_{-}) +f^*\mathcal{L}_{-} -f \mathcal{Q}^*
    & a \mathcal{L}_{+} -b (\mathcal{L}_{-}+\mathcal{L}_{+})+c \mathcal{L}_{-} +2 \mathcal{I} \Re[e]
    & d \mathcal{L}_{+} -d^* \mathcal{Q}^* -f (\mathcal{L}_{-}+\mathcal{R}_{+}) +f^*\mathcal{I} \\
  -a \mathcal{Q} +b \mathcal{I} -c \mathcal{Q}^* -e^*(\mathcal{R}_{+}^*+\mathcal{R}_{-})
  &  d^*\mathcal{L}_{+}  -d \mathcal{Q} -f^* (\mathcal{L}_{-}+\mathcal{R}_{+}^*) +f \mathcal{I}
  & b \mathcal{L}_{+} -c \mathcal{L}_{-} -2 \Re[e \mathcal{Q}] \\
 \end{bmatrix}.
\end{align}

\end{widetext}

\end{document}